\documentclass[aps,prl,twocolumn,showpacs,groupedaddress]{revtex4}  % for review and submission 

\usepackage{amsmath,amsthm,amssymb,enumerate}
\usepackage{graphicx}
\usepackage{hyperref}

\newcommand{\set}[1]{\left\{#1\right\}}
\newcommand{\bra}[1]{\left\langle#1\right\vert}
\newcommand{\ket}[1]{\left\vert#1\right\rangle}
\newcommand{\braket}[2]{\left\langle#1\middle|#2\right\rangle}
\newcommand{\ketbra}[2]{\left|#1\middle\rangle \! \middle\langle#2\right|}

\newcommand{\tr}{\text{Tr}}

\newcommand{\norm}[1]{\left\Vert#1\right\Vert}

\newcommand{\A}{\mathcal{A}}

\newcommand{\R}{\mathbb{R}}
\newcommand{\C}{\mathbb{C}}

\newcommand{\SBH}{S_\text{BH}}

\newcommand{\Hilb}{\mathcal{H}}

\newcommand{\Oc}{{\overline{\Omega}}}
\newcommand{\dO}{{\partial \Omega}}
\newcommand{\Pu}{\mathcal{P}}

\newcommand{\HS}{{\Hilb_\Sigma}}
\newcommand{\HO}{{\Hilb_\Omega}}
\newcommand{\HOc}{{\Hilb_\Oc}}
\newcommand{\HdO}{{\Hilb_\dO}}
\newcommand{\HIH}{\Hilb_\text{IH}}

\newcommand{\HOP}{\Hilb_\Omega^\Pu}

\newcommand{\HdOP}{\Hilb_\dO^\Pu}

\theoremstyle{definition}
\newtheorem*{dfn}{Definition}

\begin{document}

\title{Entanglement Entropy in Loop Quantum Gravity}
\author{William Donnelly}
\affiliation{
Department of Applied Mathematics \\
University of Waterloo \\
Waterloo, Ontario N2L 3G1, Canada
}
\email{wdonnelly@math.uwaterloo.ca}

\begin{abstract}
The entanglement entropy between quantum fields inside and outside a black hole horizon is a promising candidate for the microscopic origin of black hole entropy.
We show that the entanglement entropy may be defined in loop quantum gravity, and compute its value for spin network states. 
The entanglement entropy for an arbitrary region of space is expressed as a sum over punctures where the spin network intersects the region's boundary.
Our result agrees asymptotically with results previously obtained from the isolated horizon framework, and we give a justification for this agreement.
We conclude by proposing a new method for studying corrections to the area law and its implications for quantum corrections to the gravitational action.
\end{abstract}

% 04.60.Pp 	Loop quantum gravity, quantum geometry, spin foams
% 03.65.Ud 	Entanglement and quantum nonlocality
% 04.70.Dy 	Quantum aspects of black holes, evaporation, thermodynamics

\maketitle

\section{Introduction}

Quantum field theory in curved space indicates that black holes radiate thermally with an entropy given by the Bekenstein-Hawking formula
\begin{equation} \label{sbh}
\SBH = \frac{A}{4} \frac{k_B c^3}{G \hbar}
\end{equation}
where A is the area of the black hole event horizon \cite{bardeen-1973, hawking-1975}.
A similar entropy is present for acceleration horizons in flat space \cite{unruh-1976} and cosmological event horizons in de Sitter space \cite{gibbons-1977}.
It has therefore been suggested that every causal horizon possesses an entropy proportional to its area, so that (\ref{sbh}) represents a universal horizon entropy density \cite{jacobson-2003b}.

Although there is no universally agreed-upon source of black hole entropy, a strong candidate is the entanglement entropy of quantum fields inside and outside the black hole horizon \cite{bombelli-1986}.
This approach is based on the observation that entanglement between fields inside and outside the horizon causes any globally pure state to become mixed when restricted to the exterior of the black hole.
The entanglement entropy has the main features expected from a statistical description of black hole entropy:
it scales like the area of the horizon in the presence of a Planck-scale cutoff \cite{bombelli-1986,srednicki-1993} and under the assumption of causality it satisfies the generalized second law of thermodynamics \cite{sorkin-1986}.

In the previously studied case of a scalar field on a flat background, the entanglement entropy of a spherical region diverges in the absence of an ultraviolet cutoff.
It has been suggested that quantum gravity could act as an ultraviolet cutoff, rendering the entanglement entropy finite.
Moreover, if the gravitational field is quantized it will also contribute to the entanglement entropy.
Therefore any investigation of the relation between entanglement entropy and the Bekenstein-Hawking entropy must be done in a theory of quantum gravity.

The goal of this paper is to study the entanglement entropy of the gravitational field within the framework of loop quantum gravity.
The state of the gravitational field will be described by a spin network state \cite{rovelli-1995, ashtekar-1997b}.
The spin network states are chosen because of their interpretation as states of discrete geometry.
In particular, spin network states are eigenstates of the area operator, allowing a comparison between the entanglement entropy and the horizon area. 

Entanglement entropy has previously been computed in loop quantum gravity for black hole coherent states in spherically symmetric spacetimes with apparent horizons \cite{dasgupta-2006}.
In contrast with previous work the present result is obtained without assumption of symmetry or of particular boundary conditions at the horizon.

\section{Entanglement entropy and the Schmidt decomposition} \label{section:schmidt}

We first define the entanglement entropy and review some of its well-known properties, including its relation to the Schmidt decomposition.

Let $\mathcal{M}$ be a spacetime manifold  decomposed as $\mathcal{M} = \R \times \Sigma$ into time and space.
Let $\Hilb_\Sigma$ denote the space of wave functionals $\psi : \Phi_\Sigma \to \C$ where $\Phi_\Sigma$ is a suitably defined space of field configurations.
A partition of space $\Sigma = \Omega \cup \Oc$ gives a tensor product decomposition of the Hilbert space 
\begin{equation} \label{schmidt:tensor}
\HS = \HO \otimes \HOc
\end{equation}

Let $\ket{\psi} \in \HS$ be a pure state. 
For each region $\Omega$ there is an associated mixed state in $\HO$ given by tracing over degrees of freedom in $\Oc$,
\begin{equation} \label{schmidt:spectrum}
\rho(\Omega) = \tr_\HOc \left( \ketbra{\psi}{\psi} \right)
\end{equation}
This state encodes all information that can be obtained about $\ket{\psi}$ by performing measurements localised in $\Omega$.

The entanglement entropy of the region $\Omega$ is defined as 
\begin{equation}
S_E(\Omega) \equiv S(\rho(\Omega))
\end{equation}
where $S(\rho) = -\tr{ (\rho \log \rho)}$ is the von Neumann entropy.

\begin{dfn}[The Schmidt Decomposition]
Let $\ket{\psi} \in \HO \otimes \HOc$.
Then there exist orthonormal sets $\set{\ket{\psi^\Omega_i}} \subset \HO$ and $\set{\ket{\psi^\Oc_i}} \subset \HOc$ and positive real numbers $\set{\lambda_i}$ such that
\begin{equation}
\ket{\psi} = \sum_{i \in I} \sqrt{\lambda_i} \ket{\psi^\Omega_i} \otimes \ket{\psi^\Oc_i}
\end{equation}
The numbers $\set{\sqrt{\lambda_i}}$ are called the \emph{Schmidt coefficients}, and the number of terms in the sum is the \emph{Schmidt rank}.
\end{dfn}

From the Schmidt decomposition, we can compute the diagonal form of the reduced density matrices
\begin{align}
\rho(\Omega) &= \sum_{i \in I} \lambda_i \ket{\psi^\Omega_i}\bra{\psi^\Omega_i} \\
\rho(\Oc) &= \sum_{i \in I} \lambda_i \ket{\psi^\Oc_i}\bra{\psi^\Oc_i}
%\qquad \rho(\Oc) = \sum_{i \in I} \lambda_i \ket{\psi^\Oc_i}\bra{\psi^\Oc_i}
\end{align}
This shows that the two reduced density matrices have the same nonzero spectrum. 
It follows, as is well known, that the entanglement entropy is symmetric and can be computed from the Schmidt coefficients
\begin{equation} \label{schmidt:entropy}
S_E(\Omega) = S_E(\Oc) = -\sum_{i \in I} \lambda_i \log \lambda_i
\end{equation}
In order to compute the entanglement entropy of a spin network state, it is therefore sufficient to compute its Schmidt decomposition.

\section{Entanglement entropy of spin network states} \label{section:calculation}

% Generalized connections, Hilbert space
In loop quantum gravity, the space of fields is the space $\overline\A$ of generalized connections on $\Sigma$.
We will consider the Hilbert space $\Hilb^0_\Sigma$ of cylindrical functions, which is spanned by the extended spin network states \cite{ashtekar-1997b}.

% ESNs
An \emph{extended spin network} is a tuple $S = (\Gamma, \vec{\jmath}, \vec{J}, \vec{\imath}, \vec{m})$ where
\begin{itemize}
\item $\Gamma$ is a graph in $\Sigma$ consisting of nodes $v_1, \ldots, v_N$ and links $\gamma_1, \ldots, \gamma_L$,
\item $\vec{\jmath} = (j_1, \ldots, j_L)$ where $j_\ell \in \set{\frac{1}{2}, 1, \frac{3}{2}, \ldots}$ labels a non-trivial irreducible representation of SU(2),
\item $\vec{J} = (J_1, \ldots, J_N)$ where $J_N \in \set{0, \frac{1}{2}, 1, \frac{3}{2}, \ldots}$ labels a possibly trivial irreducible representation of SU(2),
\item $\vec{\imath} = (i_1, \ldots, i_N)$ where $i_n$ is an intertwining operator from the representations of all incoming edges and the spin $J_n$ representation to the representations of all outgoing edges, and
\item $\vec{m} = (m_1, \ldots, m_N)$ where $m_n$ is a vector in the spin $J_n$ representation space $V_{J_n} \cong \C^{2J_n + 1}$.
\end{itemize}

% ESN states
Let $A \in \overline\A$ be a generalized connection, $A(\gamma)$ the holonomy along curve $\gamma$ and $R^j(A(\gamma))$ its spin j representation.
Then the extended spin network state $\ket{S}$ is defined by 
\begin{equation} 
\braket{A}{S} = \left( \bigotimes_{\ell=1}^L R^{j_\ell} (A(\gamma_\ell))  \otimes \bigotimes_{n=1}^N m_n \right) \circ \left( \bigotimes_{n=1}^N i_n \right)
\end{equation}
Throughout we will assume that all intertwiners $i_n$ are normalized so that $\tr(i_n^*i_n) = 1$ and vectors $m_n$ are normalized so that $\norm{m_n} = \sqrt{2J_n + 1}$.

% Orthogonality
An important property of these states that we will use is their orthogonality.
Suppose that $S = (\Gamma, \vec{\jmath}, \vec{J}, \vec{\imath}, \vec{m})$ and $S' = (\Gamma, \vec{\jmath}, \vec{J}, \vec{\imath}, \vec{m}')$ so that $S$ and $S'$ differ only by their set of vectors. Then 
\begin{equation} \label{esn:orthogonal}
\braket{S}{S'} = \prod_{n=1}^N \frac{\braket{m_n}{m'_n}}{2J_n+1}
\end{equation}

% Inserting a vertex
The second important property of the extended spin network states we will use is the freedom to insert vertices.
Let $S$ be an extended spin network. 
Given an arbitrary point $v$ on the curve $\gamma_L$, we can split $\gamma_L$ into $\gamma_L = \gamma_L' \circ \gamma_L''$ so that $\gamma_L'$ and $\gamma_L''$ meet at $v$.
Let $\Gamma'$ be the graph with links $\gamma_1, \ldots, \gamma_{L-1}, \gamma_L', \gamma_L''$ and vertices $v_1, \ldots, v_N, v$.
Let $S'$ be the extended spin network $(\Gamma', \vec{\jmath}\,', \vec{J}', \vec{\imath}\,', \vec{m}')$ where $\vec{\jmath}\,' = (j_1, \ldots, j_L, j_L)$, $\vec{J}' = (J_1, \ldots, J_n, 0)$, $\vec{\imath}\,' = (i_1, \ldots, i_n, \frac{1}{\sqrt{2j+1}} I)$, $\vec{m}' = (m_1, \ldots, m_n, 1)$.
Then the resulting state $\ket{S'}$ is equivalent to $\ket{S}$, $\ket{S'} = \ket{S}$.

\subsection{Spin network states} \label{section:spinnet}

We now consider the entanglement entropy of a general spin network state.
A \emph{spin network} is an extended spin network for which the representation attached to each node is trivial, in other words $J_n = 0$, $m_n = 1$.

% Divide Gamma into two pieces 
Suppose that $\Omega$ is a subset of $\Sigma$ whose boundary $\dO$ intersects $\Gamma$ only at links.
We can insert nodes with intertwiners at all points where the graph $\Gamma$ intersects the boundary $\dO$.
Let $P$ be the number of points where $\Gamma$ intersects $\dO$, $N_\Omega$ the number of vertices of $\Gamma$ in $\Omega$, and $N_\Oc$ the number of vertices in $\Oc$.
We can partition the nodes $v_1, \ldots, v_N$ so that 
$v_n \in \dO$ for $n = 1, \ldots, P$, 
$v_n \in \Omega$ for $n = P+1, \ldots, P + N_\Omega$, 
$v_n \in \Oc$ for $n = P + N_\Omega + 1, \ldots, N$.
Similarly, the links $\gamma_1, \ldots, \gamma_L$ can be partitioned so that $\gamma_\ell \in \Omega$ for $\ell = 1, \ldots, L_\Omega$ and $\gamma_\ell \in \Oc$ for $\ell = L_\Omega + 1, \ldots, L$.

% Apply the Schmidt decomposition
By the construction for inserting vertices, $S$ has $i_p = \frac{1}{\sqrt{2\tilde{\jmath}_p+1}} I$ for $p = 1, \ldots, P$ where $\tilde{\jmath}_p$ denotes the spin of the edges incident to node $p$.
Each of these intertwiners $i_p$ may be expanded in an orthogonal basis of $V_{\tilde{\jmath}_p}$, $\set{e_1, \ldots, e_{2{\tilde{\jmath}_p}+1}}$.
\begin{equation}
i_p = \frac{1}{\sqrt{2\tilde{\jmath}_p+1}} \sum_{a_p = 1}^{2\tilde{\jmath}_p + 1} \ketbra{e_{a_p}}{e_{a_p}}
\end{equation}
We can apply this decomposition to each of the intertwiners on the boundary.
Letting $\vec{a} = (a_1, \ldots, a_P)$, 
\begin{equation} \label{spinnet:schmidt}
\ket{S} = \left( \prod_{p =1}^P \frac{1}{\sqrt{2\tilde{\jmath}_p+1}} \right) \sum_{\vec{a}} \ket{S_\Omega, \vec{a}} \otimes \ket{S_\Oc, \vec{a}}
\end{equation}
where the sum over $\vec{a}$ means to sum over all n-tuples $(a_1, \ldots, a_P)$ with $a_p = 1, \ldots, 2\tilde{\jmath}_p + 1$.

% The extended spin network state S_Omega, a 
The state $\ket{S_\Omega, \vec{a}}$ is an extended spin network state with graph $\Gamma_\Omega$ consisting of the links $\gamma_\ell, \ell = 1, \ldots, L_\Omega$ and vertices $v_n, n = 1, \ldots, P + N_\Omega$.
The labels on the links are unchanged, $j_\ell' = j_\ell$ for $\ell = 1, \ldots, L_\Omega$. The existing nodes are also unchanged $J'_n = J_n$, $i'_n = i_n$ and $m_n' = m_n$ for $n = P + 1, \ldots, P + N_\Omega$.
The inserted vertices on the boundary have $J_p' = \tilde{\jmath}_p$, $i_p' = \frac{1}{\sqrt{2\tilde{\jmath}_p+1}} I$ and $m_p' = \sqrt{2\tilde{\jmath}_p + 1} e_{a_p}$ for $p = 1, \ldots, P$.

% Schmidt rank and entanglement
By the orthogonality relation (\ref{esn:orthogonal}), these states form an orthonormal set.
Therefore (\ref{spinnet:schmidt}) is a Schmidt decomposition of $\ket{S}$. The Schmidt rank is 
\begin{equation} \label{spinnet:rank}
N = \prod_{p = 1}^P (2\tilde{\jmath}_p+1)
\end{equation}
From (\ref{schmidt:entropy}) the entanglement entropy of $\ket{S}$ is
\begin{equation} \label{spinnet:entropy}
S_E(\Omega) = \sum_{p = 1}^P \log (2\tilde{\jmath}_p + 1)
\end{equation}

% Reduced density matrix
The Schmidt decomposition also allows us to compute the reduced density matrix corresponding to the region $\Omega$,
\begin{equation}
\rho(\Omega) = \frac{1}{N} \sum_{\vec{a}} \ketbra{S_\Omega, \vec{a}}{S_\Omega, \vec{a}} \label{spinnet:mixed}
\end{equation}

Note that although the individual states $\ket{S_\Omega, \vec{a}}$ transform non-trivially under a gauge transformation, the linear combination (\ref{spinnet:mixed}) is gauge-invariant.

\subsection{Intertwiner on the boundary}

So far it has been assumed that no nodes of the spin network $S$ lie on the boundary of $\Omega$, in which case the entanglement entropy depends only the edges of $S$ that intersect the boundary.
In the case where a node lies on the boundary, the entanglement entropy depends on the intertwiner assigned to the boundary node.

In this case we can let $P$ be the number of intertwiners on the boundary
\begin{equation}
\ket{S} = \sum_{\vec{a}, \vec{b}} \left( \bigotimes_{p = 1}^P i_p \right)^{\vec{a}}_{\vec{b}} \left\vert S_\Omega,\vec{a} \middle\rangle \otimes \middle\vert S_\Oc, \vec{b} \right\rangle
\end{equation}
Here the superscript $\vec{a}$ and the subscript $\vec{b}$ index the matrix elements of the intertwiner.
Because the set of states $\ket{S_\Omega, \phantom{\vec{b}}\!\! \vec{a}}$ and $\ket{S_\Oc, \vec{b}}$ are orthogonal, this formula reduces to the entanglement entropy of the tensor product of all boundary intertwiners.

By additivity of the von Neumann entropy across tensor products, the entanglement entropy is
\begin{equation}
S_E(\Omega) = \sum_{p = 1}^P S_E(i_p)
\end{equation}
In this equation $i_p$ is viewed as an entangled state between the representations of edges incident from $\Omega$ and those incident from $\Oc$.

Note that in the case where $i_p$ is a multiple of the identity this formula reduces to $S_E(i_p) = \log(2\tilde{\jmath}_p + 1)$ as in equation (\ref{spinnet:entropy}).

\section{Relation to the isolated horizon framework} \label{section:boundary}

The expression (\ref{spinnet:rank}) for the Schmidt rank of the spin network state coincides with the dimension of the boundary Hilbert spaces defined in the isolated horizon framework \cite{ashtekar-1997a,ashtekar-2000b}.
The analogy between the isolated horizon framework and entanglement entropy was first suggested by Husain, who showed that  the calculation of black hole entropy in the isolated horizon framework does not depend on the details of the boundary conditions \cite{husain-1998}.
To elucidate the relationship between these two theories it is necessary to first introduce the relevant aspects of the isolated horizon framework.

In the isolated horizon framework, the horizon is treated as an inner boundary of space obeying the isolated horizon boundary conditions.
The gravitational action acquires a surface term proportional to the action of Chern-Simons theory.
The states are therefore elements of $\HO \otimes \HdO$ where $\HO$ is the ``bulk space'' of cylindrical functions on $\Omega$ and $\HdO$ is the ``boundary space'' of $U(1)$ Chern-Simons theory on $\partial \Omega$.

The isolated horizon boundary conditions are implemented as an operator equation on $\HO \otimes \HdO$ restricting states to a subspace
\begin{equation} \label{boundary:hih}
\HIH = \bigoplus_\Pu \HOP \otimes \HdOP / \text{Gauge}
\end{equation}
Where $\Pu$ runs over all finite set of punctures labelled by spins $\tilde{\jmath}_p$.
The space $\HOP$ is spanned by the open spin network states $\ket{S_\Omega, \vec{a}}$, where $S_\Omega$ runs over all extended spin networks intersecting $\dO$ in the set of labelled punctures $\Pu$.
$\HdOP$ is the space of Chern-Simons states on the punctured surface $\partial \Omega - \Pu$.

A partial trace over the bulk space $\HO$ is performed, yielding a maximally mixed state on $\HdOP$. 
The entanglement entropy is therefore given by $\log \dim \HdOP$, where in the limit of a large number of punctures $\log \dim \HdOP$ is given by 
\begin{equation} \label{boundary:dimension}
\dim \HdOP \sim \prod_{p = 1}^P (2 \tilde{\jmath}_p + 1)
\end{equation}
which is the same as the Schmidt rank of a spin network that intersects $\partial \Omega$ at the points $\Pu$.

To see why these quantities should agree, consider an arbitrary boundary theory with Hilbert space $\Hilb_\Omega \otimes \Hilb_\dO$.
We will make the assumption that the boundary conditions should not affect physics in $\Omega$.
Therefore for each spin network state $\ket{S} \in \HO \otimes \HOc$ there should exist a state $\ket{S'} \in \Hilb_\Omega \otimes \Hilb_\dO$ that describes the same physics on the exterior
\begin{equation} \label{boundary:trace}
\tr_\HOc(\ketbra{S}{S}) = \tr_\HdO(\ketbra{S'}{S'})
\end{equation}

Now consider the mixed state of the boundary $\rho_\dO = \tr_\HO(\ketbra{S'}{S'})$.
By equation (\ref{schmidt:spectrum}), $\rho_\dO$ has the same nonzero spectrum as $\rho_\Omega$.
Therefore the range of $\rho_\dO$ is a subspace whose dimension is given by the Schmidt rank (\ref{spinnet:rank}).
Thus equation (\ref{boundary:dimension}) is a consequence of the fact that the rank of $\rho_\Omega$ is the same as the rank of $\rho_\dO$.

By imposing the requirement that the boundary conditions should not restrict the exterior Hilbert space, we have derived the relationship between the Schmidt rank of a spin network state and the dimension of the boundary Hilbert space.
However equation (\ref{boundary:dimension}) only holds asymptotically; the dimension of $\HdOP$ is less than the Schmidt rank.

To understand the reason for disagreement, we note that a basis of the space $\HdOP$ is labelled by sequences of half-integers $\set{m_p}_{p = 1}^P$ 
satisfying $m_p \in \set{ - \tilde{\jmath}_p, -\tilde{\jmath}_p+1, \ldots, \tilde{\jmath}_p}$ and the additional \emph{spin projection constraint}
\begin{equation} \label{boundary:projection}
\sum_{p = 1}^P m_p = 0
\end{equation}
Therefore the difference between the dimension and the Schmidt rank is entirely due to the spin projection constraint.

\section{Outlook} \label{section:conclusion}

We have computed the entanglement entropy of the gravitational field in loop quantum gravity for an arbitrary region of space $\Omega$ and spin network state $\ket{S}$.
The entanglement entropy is a finite and extensive quantity that depends linearly on the number of punctures of the horizon.
No assumptions have been made about the region $\Omega$, so our result applies to all causal horizons.

We have found an interesting relation between the entanglement entropy and the isolated horizon framework.
What remains to be understood is the mismatch between the two theories due to the spin projection constraint.
It would therefore be of interest to see whether this constraint is necessary, or whether it can be relaxed by a less restrictive choice of boundary conditions.
Of particular interest is the relation between the spin projection constraint and the requirement that $\dO$ have spherical topology.

Finally, the fact that the entropy is extensive over the horizon allows thermodynamics to be applied locally to horizons in loop quantum gravity \cite{jacobson-1995,eling-2006}.
It has been shown that under the assumption that the horizon entropy is extensive, properties of the effective action for gravity can be computed from the entropy density.
However, in order to apply these arguments it is necessary to express the entropy as a function of geometric variables instead of spin networks.
This suggests the question of whether there exists a geometric quantity $Q$ such that
\begin{equation}
\left( \widehat{\oint_\dO Q} \right) \ket{S} = \sum_{p = 1}^P \log (2 \tilde{\jmath}_p + 1) \ket {S}
\end{equation}
Results in the context of classical general relativity suggest that $Q$ is related to a Noether charge of the gravitational effective action \cite{wald-1993}.
Together, these results suggest that horizon thermodynamics could be a powerful tool for studying quantum corrections to the gravitational action in loop quantum gravity.

\bibliographystyle{prsty}

\end{document}